\documentclass{elsart}
\usepackage{epsfig}

\begin{document}

\begin{frontmatter}

\title{Charm-anticharm baryon production asymmetries in photon-nucleon 
interactions}
\collab{The FOCUS Collaboration$^\star$}

\author[ucd]{J.~M.~Link}
\author[ucd]{P.~M.~Yager}
\author[cbpf]{J.~C.~Anjos}
\author[cbpf]{I.~Bediaga}
\author[cbpf]{C.~G\"obel}
\author[cbpf]{A.~A.~Machado}
\author[cbpf]{J.~Magnin}
\author[cbpf]{A.~Massafferri}
\author[cbpf]{J.~M.~de~Miranda}
\author[cbpf]{I.~M.~Pepe}
% ...

\address[ucd]{University of California, Davis, CA 95616} 
\address[cbpf]{Centro Brasileiro de Pesquisas F\'\i sicas, Rio de Janeiro, RJ, Brasil} 
% ...

\endnote{\small $^\star$ See http://www-focus.fnal.gov/authors.html for
additional author information}

\begin{abstract}
We report measurements of the charm-anticharm production asymmetries for
$\Lambda_c^+$, $\Sigma_c^{++}$, $\Sigma_c^0$, $\Sigma_c^{++*}$, 
$\Sigma_c^{0*}$, and $\Lambda_c^+(2625)$ baryons 
from the
Fermilab photoproduction experiment FOCUS (E831). 
These asymmetries are integrated over
the region where the spectrometer has good acceptance. In addition,
we have obtained results for the photoproduction asymmetries of the
$\Lambda_c$ baryons as functions of $p_L$, $p_T^2$, and $x_F$. 
The integrated asymmetry for $\Lambda_c^+$ production, 
$(\sigma_{\Lambda_c^+} \,-\, \sigma_{\Lambda_c^-})\, /\, (\sigma_{\Lambda_c^+} \,+\, \sigma_{\Lambda_c^-})$,
is $0.111 \pm 0.018 \pm 0.012$, significantly different from zero.  The asymmetries
of the excited states are consistent with the $\Lambda_c$ asymmetry.
\end{abstract}

\end{frontmatter}


The FOCUS experiment uses a photon beam on a Beryllium Oxide target to produce charm
particles.  In high energy photon-hadron interactions, pairs of charm-anticharm quarks
are produced predominantly through photon-gluon fusion~\cite{lo}. At leading order
in Quantum Chromodynamics (QCD), the produced charm and anticharm particles are identically
distributed in the kinematic variables.
At next-to-leading order, small
asymmetries are expected between charm and anticharm production~\cite{nlo1,nlo2,nlo3}. However,
these predicted asymmetries would be too small to be measured at the current level of 
experimental statistics.
Interactions between the produced charm quarks and those in the struck nucleon during
hadronization can also induce production asymmetries.  One common model is that
of string fragmentation as implemented in \textsc{Pythia}~\cite{pythia}.  In this model,
the charm and anticharm quarks are connected through color strings to the quarks in the 
struck nucleon. 
The energy in these strings is converted to particles by ``popping'' $q\bar{q}$ pairs out
of the vacuum.  The simplest asymmetry example occurs when a charm quark is connected to
the diquark in the nucleon by a low energy string with insufficient energy to produce 
any additional particles.
In this case, the charm quark can combine with valence $u$ and $d$ quarks to form 
a $\Lambda_c$ while a $\bar{c}$ quark can only form mesons when it combines with the
valence quarks.

In this letter we present a high-statistics measurement of the production asymmetry
of $\Lambda_c$ baryons from photon-nucleon interactions,
providing the first convincing evidence for a non-zero asymmetry.
This letter also contains the first measurements of this asymmetry as functions of
$p_L$, $p_T^2$, and $x_F$.  In addition, the production asymmetries of the excited
charm baryons $\Sigma_c^{++}$, $\Sigma_c^0$, $\Sigma_c^{++*}$, $\Sigma_c^{0*}$, and
$\Lambda_c^+(2625)$ are presented for the first time. All of these states decay to 
a $\Lambda_c$ plus one or two charged pions.

The FOCUS (Fermilab E831) experiment was designed to study charm particle
physics. Charmed hadrons are produced by the interaction of high energy
photons  ($\left<E\right>\approx 175$ GeV for events in which a charm decay was reconstructed)
 with a segmented Beryllium
Oxide (BeO) target.  The photons are produced by bremsstrahlung in a lead
target with a 
300 GeV $e^+/e^-$ beam.  Vertex reconstruction is performed using
four silicon strip planes interleaved with segments of the target followed by
a 12 plane silicon strip vertex detector.  Downstream of the
vertex detector, tracking and momentum measurements are made using five
stations of multiwire proportional chambers and two large aperture magnets
with opposite polarity. Three multicell \v{C}erenkov counters
operating in threshold mode are used to identify electrons,
pions, kaons, and protons over a wide range of momenta. The spectrometer
also contains a hadron calorimeter, two electromagnetic calorimeters,
and two muon detectors.  Data were collected during the 1996--97 fixed-target
run.


The $\Lambda_c^+$ particles are reconstructed using the $pK^-\pi^+$ decay 
mode.\footnote{Charge conjugate states are implied, unless stated  otherwise}  
The decay vertex is formed from
three charged tracks in the event.  The vector sum of their momenta is
projected back toward the target and used as a {\it seed} to intersect
with at least one other track in the event to form a production
vertex.  We require the production and decay vertices to be separated
by at least 5.5 $\sigma_\ell$, where $\sigma_\ell$ is the uncertainty in the measured 
vertex separation.  We also place goodness-of-fit criteria on the 
primary and secondary vertices, requiring that the confidence level for
each vertex be greater than 1\%.  Identification of the decay tracks by
particle type is performed using the \v{C}erenkov detector system~\cite{cer}.  
A $\chi^2$-like variable is formed using the on/off status of all cells
within a particle's \v{C}erenkov cone ($\beta$=1). For each of the four
possibilities, electron, pion, kaon, and proton, we calculate $W_i=-2\
\Sigma_j^{\textrm{cells}}\log P_j$, where $i$ is the particle type and $j$ the cell
number.  $P_j$ is the probability that the $j^\textrm{th}$ cell will yield the
observed response given particle type $i$.  Identification is then based on
differences in the $W_i$.
The proton candidate is required to
have the proton hypothesis favored over the kaon and pion hypotheses by 1
and 4 units of log likelihood, respectively. The kaon hypothesis for the
kaon candidate is required to be favored over the pion hypothesis by 3 such
units.  For the pion candidate, the pion hypothesis cannot be
disfavored by more than 6 units of log likelihood relative to the most
likely hypothesis. This very loose 
requirement is also applied to the pions from the excited
charm decays, described below.
To remove longer lived charm backgrounds,
the lifetime of the $\Lambda_c$ in its rest frame must be shorter than 8 times 
the world average lifetime~\cite{pdg}.
$\Lambda_c$ candidates are identified as all those events which satisfy
the above criteria and which fall into the mass range between 2.10 and
2.45~GeV/$c^2$.  
Finally, we restrict our sample to be in the
(large) phase space region given by $40<p_L<200$~GeV/$c$ and $p_T^2<6.0$~(GeV/$c)^2$.

The $\Sigma_c^{++}$, $\Sigma_c^0$, $\Sigma_c^{++*}$, and $\Sigma_c^{0*}$
candidates are reconstructed by combining the $\Lambda_c^+$ candidates
within $2\sigma$ of the mean $\Lambda_c^+$ mass with a single charged pion
track. All possible combinations in the event are tried. The confidence
level of the $\Sigma_c$ decay vertex is required to be greater than 1\%. 
To reduce systematic errors coming from the reconstruction of the
$\Lambda_c^+$ and obtain better signal-to-noise, 
we study the $\Sigma_c$ states using the difference in the
invariant mass of each $\Sigma_c$ state and the invariant mass of the
$\Lambda_c^+$, $\Delta M$.   Because the pion is typically
of low momentum, it suffers from a considerable amount of multiple
scattering, and the uncertainty on its momentum dominates the error on the
$\Sigma_c$ invariant mass.  To improve the momentum measurement,
the primary vertex is refit without this {\it soft} pion track (if possible),
and the pion direction is recomputed, forcing it to come from the refit
primary.

$\Lambda_c^+$(2625) candidates are reconstructed by combining the 
$\Lambda_c^+$ candidates within $2\sigma$ of the mean $\Lambda_c^+$ mass
with all combinations of two pions of opposite charge in the event. 
As for the $\Sigma_c$ states, we use the mass
difference plots and force the two pions to come from the primary.

FOCUS data were taken with three different beam energies and with two different 
radiators. Since the asymmetries can depend on the energy spectrum of the incident beam
photons, we include only those events in our sample
that come from
data taken with a 300 GeV $e^-/e^+$ beam on a 
lead radiator of 20\% of a radiation length.
This selection results in a charm baryon yield which
is about 75\% of the yield found using all of the data.
For reconstructed charm events passing the trigger,
the beam energy is reasonably well
described by a Gaussian with mean of 175~GeV and width
of 45~GeV.


From studies of high statistics meson decays we find no evidence that there is any 
asymmetry introduced by charge bias in the spectrometer.
Even so, since the acceptance depends on the longitudinal and
transverse momenta of the produced baryon state, if these spectra are
different for particle and antiparticle over the range for which the
asymmetry is measured, there will be an acceptance difference for the two
samples which can be corrected.  This corrected asymmetry gives the asymmetry
for an experiment with flat acceptance in $p_L$ and $p_T^2$ 
over the range $40<p_L<200$~GeV/$c$ and $p_T^2<6.0$~(GeV/$c)^2$.

To compensate for this acceptance difference as a function of $p_T^2$ and
$p_L$, the production asymmetry $A$ is determined by:

\begin{equation}
A = \frac{N/\epsilon - \overline{N}/\overline{\epsilon}}{N/\epsilon + \overline{N}/\overline{\epsilon}}
\label{eq:acorrected}
\end{equation}
\noindent where $N$ and $\overline{N}$ are the numbers of reconstructed baryons and antibaryons, 
respectively and $\epsilon(\overline{\epsilon})$ is the baryon (antibaryon) reconstruction efficiency.
The efficiencies $\epsilon$ and $\overline{\epsilon}$ are calculated using the
FOCUS Monte Carlo simulation program.  The detector simulation uses a detailed
description of the FOCUS detector.  The physics processes are generated using 
\textsc{Pythia} (version 6.127), with many \textsc{Pythia} parameters tuned to match the
observed FOCUS physics distributions.  The remaining discrepancy between data
and Monte Carlo is removed by weighting Monte Carlo events to exactly match the observed
data $p_L$ and $p_T^2$ distributions.  For the excited charm states, the particle/antiparticle
efficiencies are used to correct the asymmetry.  For the higher statistics $\Lambda_c$ decays
a further step is taken.
Since very little $p_T^2$ dependence on efficiency 
is observed, the Monte Carlo events are used to obtain the efficiency variation versus
$p_L$.  This efficiency is fit to a function which is then used to weight each event.

\begin{figure}  
  \centerline{\epsfig{figure=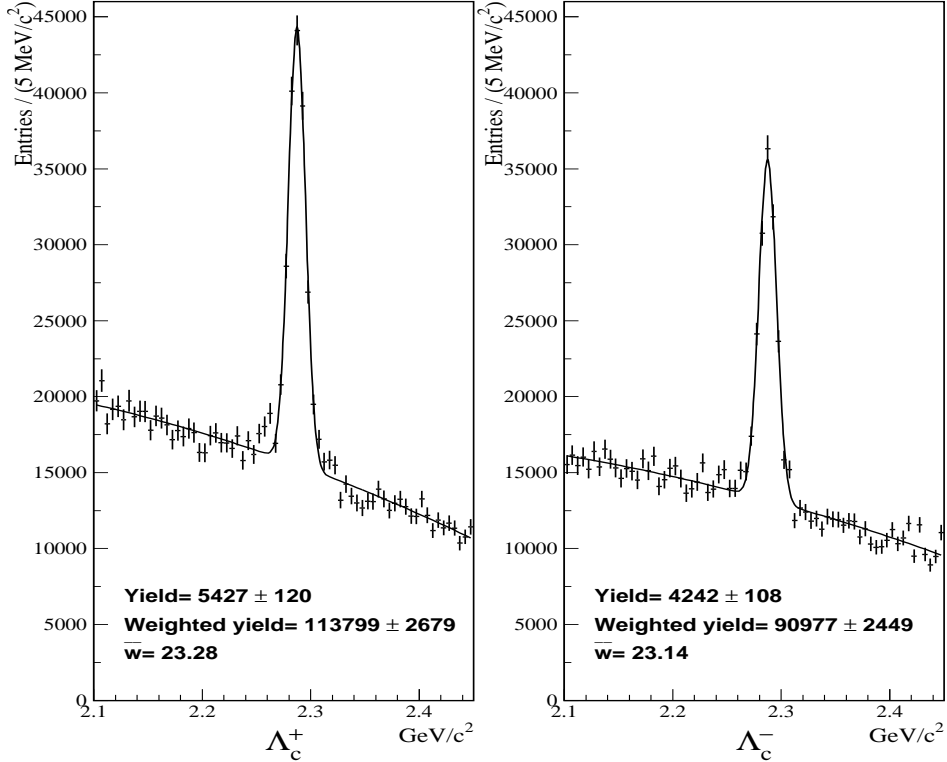,height=4.0in,width=5.0in}}
  \caption{$pK^-\pi^+$ weighted invariant mass distribution split by charge. The weights are used to account 
  for efficiency loss versus momentum on an event-by-event basis.  The average weight 
  for the whole histogram is given in the figure 
  as $\overline{w}$. The unweighted yields are also shown.
  \label{lc_signal}}
\end{figure}

The global $\Lambda_c$ asymmetry is obtained by fitting each of the two
weighted invariant mass data distributions shown in Fig.~\ref{lc_signal} with a Gaussian
signal and quadratic background.  The Gaussian mean and width are allowed
to float separately for $\Lambda_c^+$ and $\Lambda_c^-$.
The unweighted yields for 
$\Lambda_c^+$ and $\Lambda_c^-$ are 5427$\pm$120 and 4242$\pm$108,
respectively.

\begin{figure} 
  \centerline{\epsfig{figure=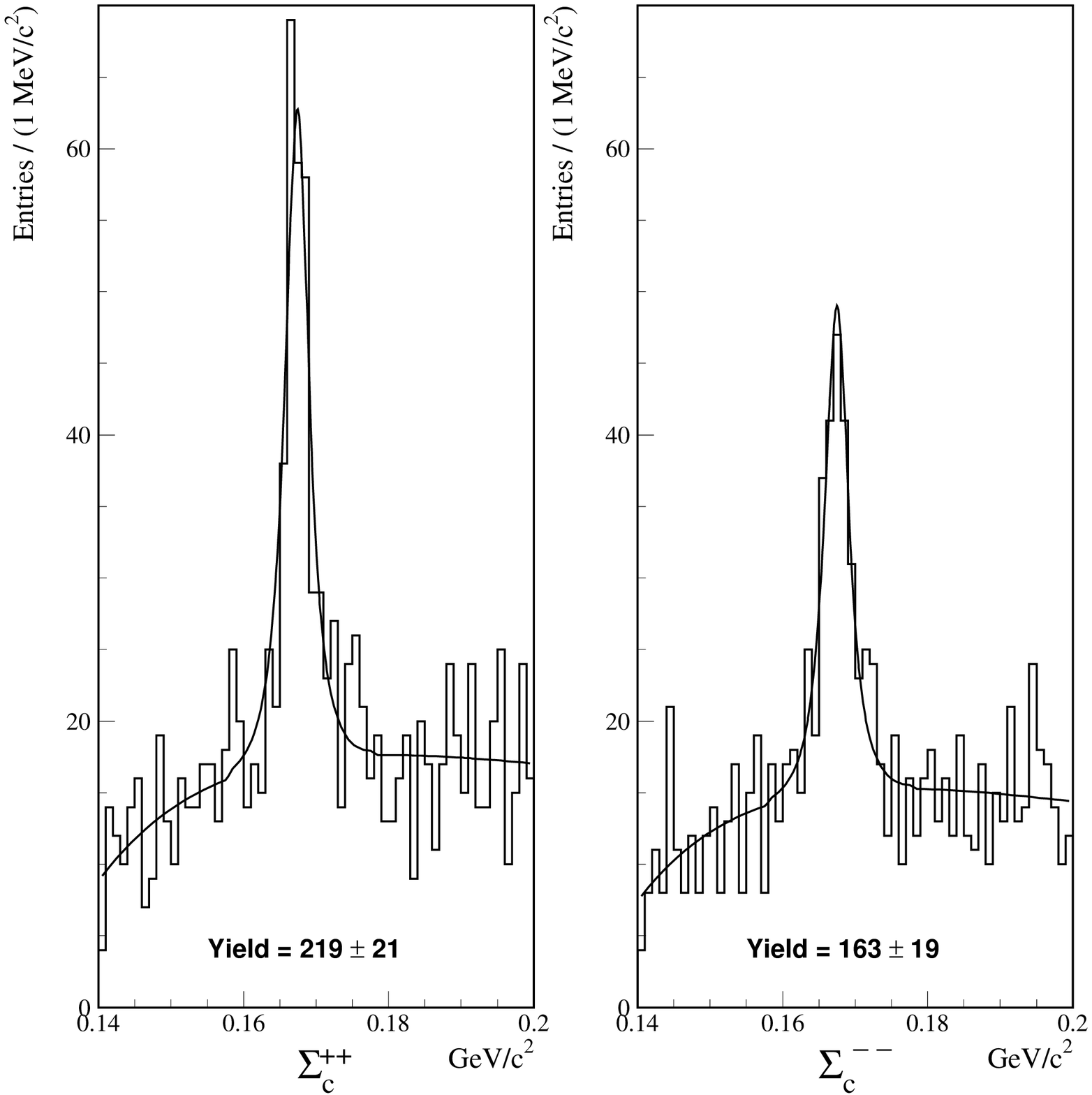,height=4.0in,width=5.0in}}  
  \caption{$\Lambda_c^+\pi^+$ invariant mass distribution split by charge.
  \label{scpp_signal}}
\end{figure}

\begin{figure} 
  \centerline{\epsfig{figure=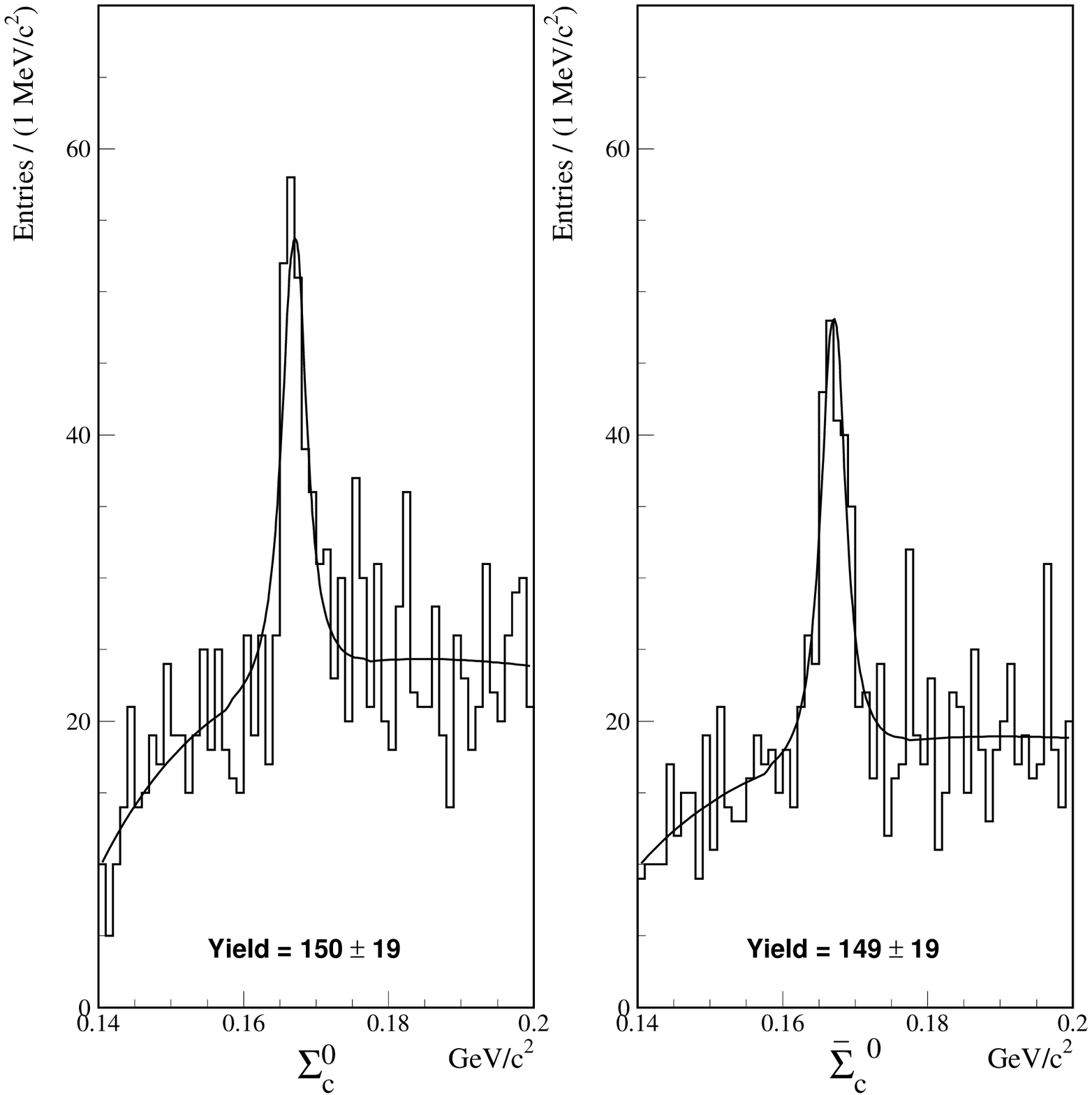,height=4.0in,width=5.0in}}
  \caption{$\Lambda_c^+\pi^-$ invariant mass distribution split by charge.
  \label{sc0_signal}}
\end{figure}

\begin{figure} 
  \centerline{\epsfig{figure=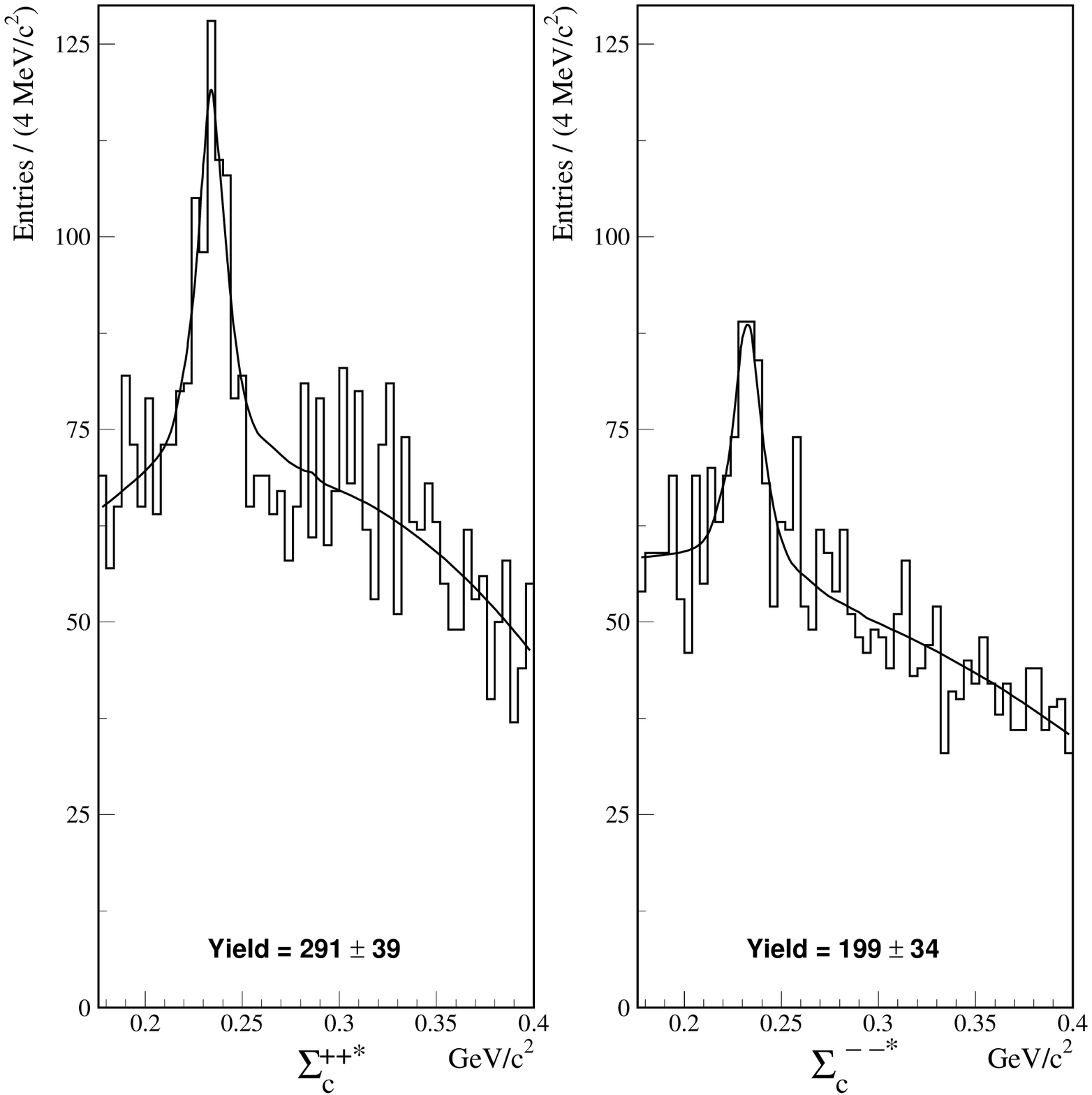,height=4.0in,width=5.0in}}
  \caption{$\Lambda_c^+\pi^+$ invariant mass distribution split by charge.
  \label{scpps_signal}}
\end{figure}

\begin{figure}   
  \centerline{\epsfig{figure=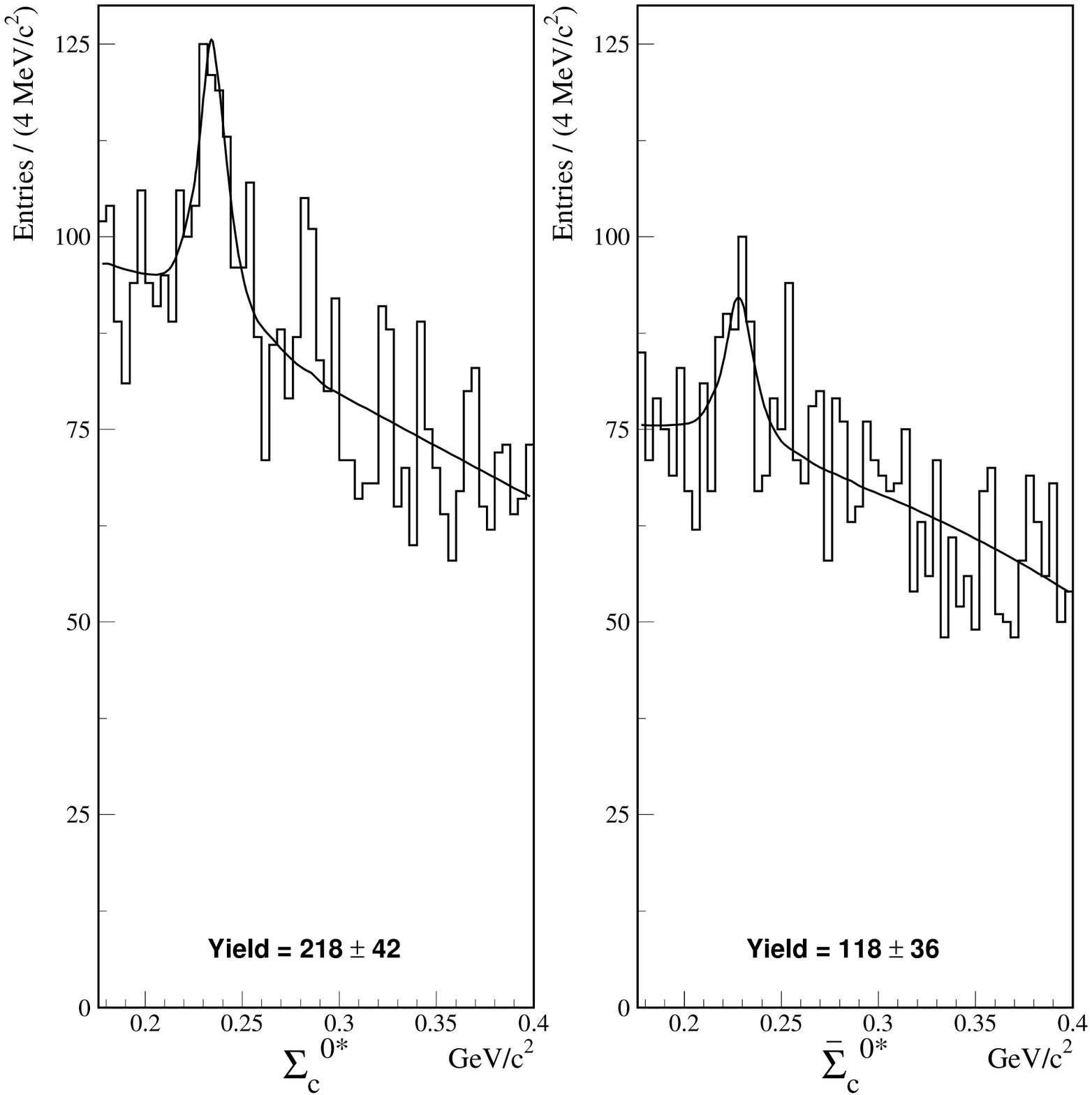,height=4.0in,width=5.0in}}
  \caption{$\Lambda_c^+\pi^-$ invariant mass distribution split by charge.
  \label{sc0s_signal}}
\end{figure}

\begin{figure} 
  \centerline{\epsfig{figure=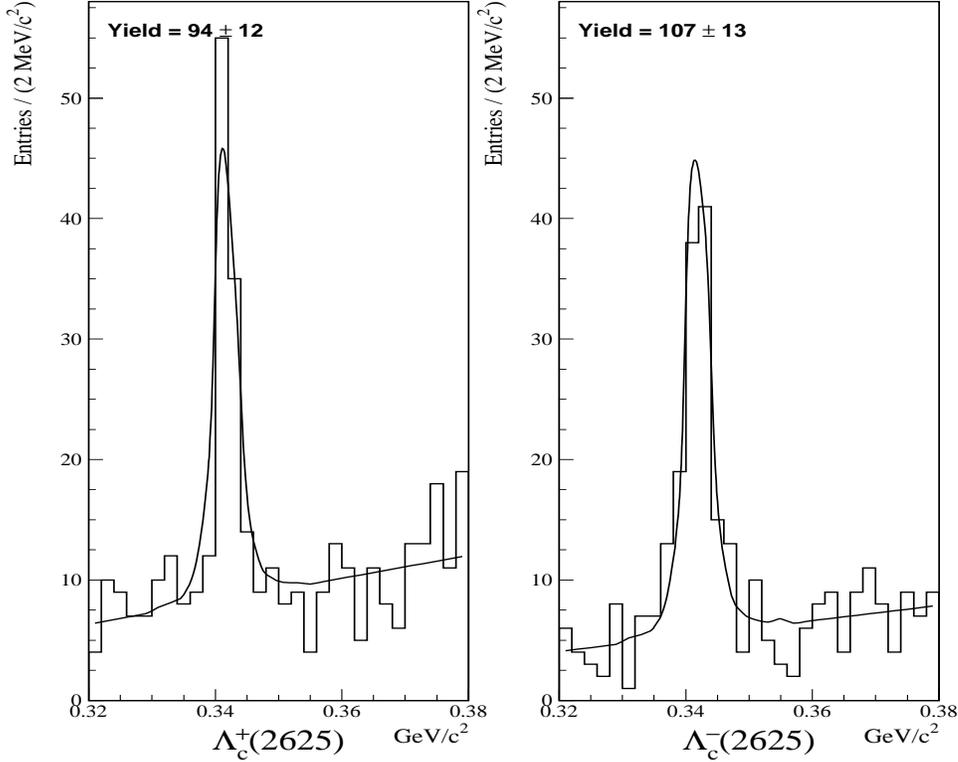,height=4.0in,width=5.0in}}
  \caption{$\Lambda_c^+\pi^+\pi^-$ invariant mass distribution split by charge.
  \label{lcs_signal}}
\end{figure}

To account for natural widths and changing resolutions, the excited charm 
baryon states are fit slightly differently.  The Monte Carlo is used
to generate the signal shape which is fit to a spline function.  This spline function,
properly normalized, is used as the signal shape for the data, with the
mass allowed to float.  The advantage in the
case of the $\Sigma_c$, and $\Sigma_c^*$ is that the spline function is able to
account for the detector resolution and natural width of the state.  For
the $\Lambda_c^*$ state, the spline function is a better model of the detector 
resolution than a Gaussian due to the small phase space.  
The background shapes for 
$\Sigma_c$, $\Sigma_c^*$ and $\Lambda_c(2625)$ are a threshold function 
$N(1 + \alpha(\Delta M - m_{\pi})\Delta M^{\beta})$, quadratic polynomial,
and linear polynomial, respectively.
The fits to each data sample are shown on the
corresponding data plots in 
Figs~\ref{scpp_signal}--\ref{lcs_signal}.
The global asymmetries are calculated from the returned yields.

\begin{figure}  
  \centerline{\epsfig{figure=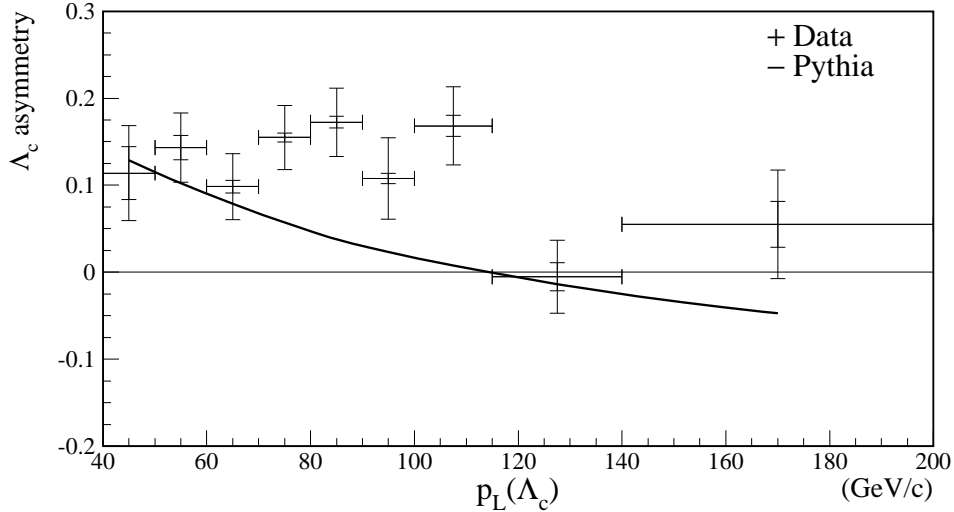,width=5.0in}}
  \caption{$p_L$ asymmetry distribution for $\Lambda_c$ along with statistical and systematic errors
  and the \textsc{Pythia} prediction -- the systematic errors are superposed on the statistical errors and
  are smaller than the statistical errors in every bin.  
  \label{asym_pz}}
\end{figure}

\begin{figure}  
  \centerline{\epsfig{figure=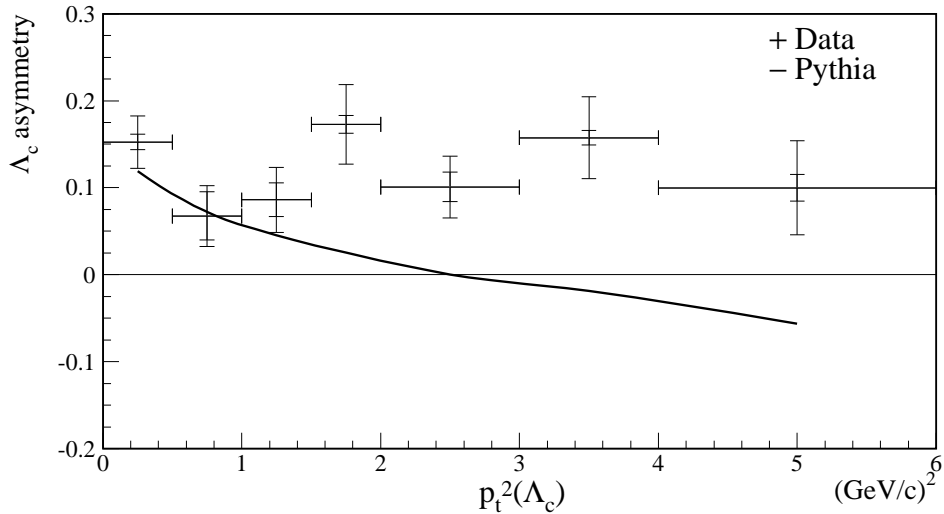,width=5.0in}}
  \caption{$p_T^2$ asymmetry distribution for $\Lambda_c$ along with statistical and systematic errors
  and the \textsc{Pythia} prediction -- the systematic errors are superposed on the statistical errors and
  are smaller than the statistical errors in every bin.  
  \label{asym_pt2}}
\end{figure}

\begin{figure}  
  \centerline{\epsfig{figure=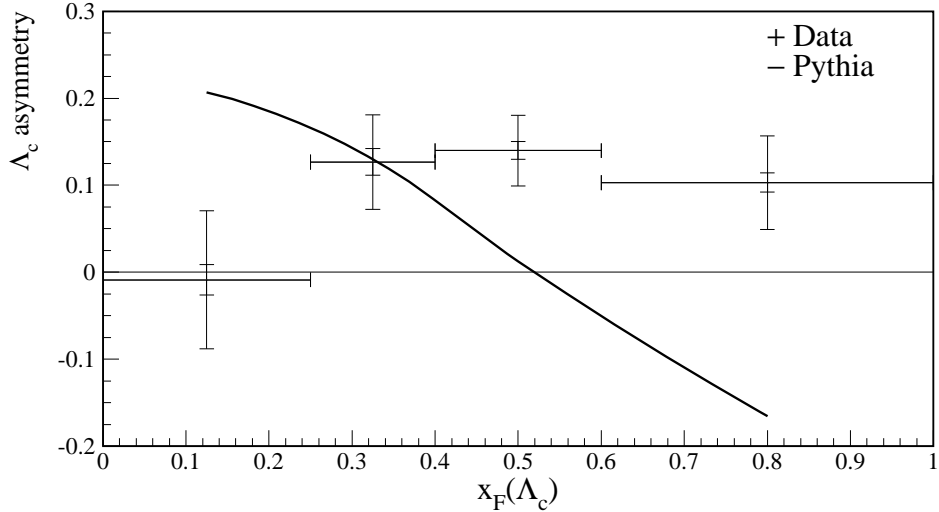,width=5.0in}}
  \caption{$x_F$ asymmetry distribution for $\Lambda_c$ along with statistical and systematic errors
  and the \textsc{Pythia} prediction -- the systematic errors are superposed on the statistical errors and
  are smaller than the statistical errors in every bin.  
  \label{asym_xf}}
\end{figure}

The $\Lambda_c$ data samples are divided into bins of each of the
kinematic variables $p_L$, $p_T^2$, and $x_F$, integrating over the full range of
the other variables.  The mass plots for each bin are fit using a Gaussian
signal and quadratic background, as for the full data set. For these fits,
the $\Lambda_c^+$ ($\Lambda_c^-$) mass is fixed to the mass obtained by
fitting the full $\Lambda_c^+$ ($\Lambda_c^-$) sample. The widths
are fixed to follow the Monte Carlo widths for each bin.
The yields from these fits are used to obtain the production asymmetry versus
$p_L$, $p_T^2$, and $x_F$.  The efficiency corrected asymmetry distributions 
for $\Lambda_c$ vs $p_L$, $p_T^2$, and $x_F$ are shown in  
Figs.~\ref{asym_pz}, \ref{asym_pt2}, and \ref{asym_xf}.  The $x_F$ measurement
requires knowledge of the incoming beam energy which is only available in about
30\% of the data sample and is therefore of lower statistics.


In our first systematic error check, background studies were performed 
to assure that feedthrough from 
other charm states does not contribute to the asymmetry.
The systematic errors are obtained by making the same measurements under different analysis
conditions.  We performed these studies for the global asymmetry and the asymmetry
versus the kinematic variables reported.

For the $\Lambda_c$ and the three excited states ($\Sigma_c$, $\Sigma_c^*$, and
$\Lambda_c^*$), fits with different bin widths and with
bins shifted by 1/2 bin were performed.
A variation which fixed the mass for baryons and antibaryons
to the value from the total sample was also studied.
We also investigated differences due to the fit functions used. 
For the $\Sigma_c$ states a Gaussian function 
with a quadratic background was utilized to fit the signals.
For the $\Sigma_c^*$ mass distributions we used a linear and cubic
fit for the background instead of the quadratic background.
For $\Sigma_c$
and $\Sigma_c^*$, fits with spline functions from two additional Monte Carlo
samples where the natural widths were varied by $\pm 1 \sigma$ were used to
estimate the uncertainty in the knowledge of the natural widths.
For the $\Lambda_c^*$ we used two different fit functions 
to provide a systematic check: a Gaussian for the signal shape 
and a linear background and the $\Lambda_c^*$ spline function with a quadratic 
background.
Additionally, for all the states we used the respective raw 
asymmetries to check for systematic problems with the efficiency correction.
For the $\Lambda_c$ we also calculated the asymmetry using the efficiencies
directly from the Monte Carlo instead of using the Monte Carlo to obtain an
efficiency function.

The r.m.s. of all of these variations is used as an estimate of the systematic error.

The global asymmetries for all of the charm baryons studied in this analysis
are shown in Table~\ref{table:asym}. The first errors are statistical and the
second are systematic.
Table~\ref{table:asym} also shows comparisons
to \textsc{Pythia} asymmetries calculated with unreconstructed events.
The \textsc{Pythia} predictions come from running version 6.203
with all parameters left at the default setting.  
The events are generated with the
correct beam energy distribution and only candidates within the nominal phase space,
$40<p_L<200$~GeV/$c$ and $p_T^2<6.0$~(GeV/$c)^2$,
are used in determining the asymmetry.
The E691 and E687 photoproduction experiments have previously reported global asymmetries for $\Lambda_c$
of $0.110 \pm 0.089$~\cite{e691} and $0.035\pm 0.076$~\cite{e687}, respectively. 
Our results are similar to those obtained by these experiments although comparisons
are not straightforward since all three experiments have different phase space
and beam energy distributions. Table~\ref{table:asym} also shows the efficiency
ratios of particles to antiparticles for the charm baryons studied.

\begin{table} [h!]
\begin{center}
\caption{Raw and corrected global production asymmetry for the charm baryons compared 
to the predictions of default \textsc{Pythia} version 6.203. The efficiency ratio of 
particles to antiparticles is also shown.}
\label{table:asym}
\begin{tabular}{lrrcr} \hline\hline
Baryon & Raw Asymmetry & Corrected Asymmetry & \textsc{Pythia} & Efficiency Ratio \\ 
\hline
$\Lambda_c^+$                & $0.123\pm 0.017$ &  $0.111\pm 0.018\pm 0.012$ & $0.073$ & $1.023\pm 0.005$\\
$\Sigma_c^{++}$              & $0.147\pm 0.073$ &  $0.136\pm 0.073\pm 0.036$ & $0.126$ & $1.024\pm 0.010$\\
$\Sigma_c^{0}$               & $0.005\pm 0.089$ &  $0.005\pm 0.089\pm 0.024$ & $0.128$ & $1.000\pm 0.009$\\
$\Sigma_c^{++\,*}$           & $0.188\pm 0.105$ &  $0.181\pm 0.105\pm 0.033$ & $0.133$ & $1.016\pm 0.006$\\
$\Sigma_c^{0\,*}$            & $0.299\pm 0.165$ &  $0.298\pm 0.165\pm 0.023$ & $0.132$ & $1.002\pm 0.006$\\
$\Lambda_c^{+}(2625)\!\!\!$  & $-0.066\pm 0.086$ & $-0.075\pm 0.087\pm 0.021$ & $N/A$  & $1.019\pm 0.017$\\
\hline\hline
\end{tabular}
\end{center}
\end{table}


We have studied the photoproduction asymmetry of $\Lambda_c^+$ versus
$\Lambda_c^-$ using the decay channel $pK^-\pi^+$.  From
$\sim$ 10,000 $\Lambda_c$ events we present the first results of this 
asymmetry as functions of $p_L$, $p_T^2$, and $x_F$.  These
results show a clear positive asymmetry. The global $\Lambda_c^+$ asymmetry 
is measured to be $0.111 \pm 0.018\pm 0.012$.

The production asymmetry of excited charm states which decay to $\Lambda_c^+$,
including the  $\Sigma_c^{++}$, $\Sigma_c^0$, $\Sigma_c^{++*}$,
$\Sigma_c^{0*}$, and $\Lambda_c^+(2625)$ was also measured for the first time. 
The measurements generally indicate a positive 
asymmetry similar to the $\Lambda_c$.  Because of the smaller sample size, however,
they are also consistent with zero. 

We find that the string fragmentation model as implemented in \textsc{Pythia} 
version 6.203 does not describe the $\Lambda_c^+$ asymmetry dependence on 
$p_L$, $p_T^2$, or $x_F$.  Our measurements indicate that the asymmetry 
shows no significant dependence in these variables.

We acknowledge the assistance of the staffs of Fermi National
Accelerator Laboratory, the INFN of Italy, 
the physics departments
of the collaborating institutions and
the Instituto de F\'{\i}sica ``Luis Rivera Terrazas'' de la
Benem\'erita Universidad Aut\'onoma de Puebla (IFUAP), M\'exico. 
This research was supported in part by the
U.~S. National Science Foundation, the U.~S. Department of Energy, the Italian
Istituto Nazionale di Fisica Nucleare and
Ministero della Istruzione, Universit\`a e
Ricerca, Organizaci\'on de los Estados Americanos (OEA), 
IFUAP-M\'exico, CONACyT-M\'exico,
the Brazilian Conselho Nacional de
Desenvolvimento Cient\'{\i}fico e Tecnol\'ogico, the Korean
Ministry of Education, and the Korean Science and Engineering Foundation.


\end{document}